\documentclass[a4paper,twocolumn,superscriptaddress,11pt,accepted=2019-03-26]{quantumarticle}
\pdfoutput=1

 \usepackage[english]{babel}
 \usepackage[T1]{fontenc}
\usepackage{amsmath}
\usepackage{hyperref}

\usepackage{natbib,hyperref}

\usepackage{tikz}
\usepackage{lipsum}

\usepackage{url}

\bibstyle{plain}

\begin{document}

\title{Analysing correlated noise on the surface code using adaptive decoding algorithms}

\author{Naomi H. Nickerson}
\affiliation{Quantum Optics and Laser Science, Blackett Laboratory, Imperial College London, Prince Consort Road, London SW7 2AZ, United Kingdom}

\author{Benjamin J. Brown}
\affiliation{Niels Bohr International Academy, Niels Bohr Institute, Blegdamsvej 17, 2100 Copenhagen, Denmark}
\affiliation{Centre for Engineered Quantum Systems, School of Physics, University of Sydney, Sydney, New South Wales 2006, Australia}
\date{\today}

\maketitle

\begin{abstract}
Laboratory hardware is rapidly progressing towards a state where quantum error-correcting codes can be realised. As such, we must learn how to deal with the complex nature of the noise that may occur in real physical systems. Single qubit Pauli errors are commonly used to study the behaviour of error-correcting codes, but in general we might expect the environment to introduce correlated errors to a system. Given some knowledge of structures that errors commonly take, it may be possible to adapt the error-correction procedure to compensate for this noise, but performing full state tomography on a physical system to analyse this structure quickly becomes impossible as the size increases beyond a few qubits. Here we develop and test new methods to analyse blue a particular class of spatially correlated errors by making use of parametrised families of decoding algorithms. We demonstrate our method numerically using a diffusive noise model. We show that information can be learnt about the parameters of the noise model, and additionally that the logical error rates can be improved. We conclude by discussing how our method could be utilised in a practical setting blue and propose extensions of our work to study more general error models.
\end{abstract}

\section{Introduction}

The success of scalable quantum computation depends on our ability to diagnose and repair errors incident to the underlying physical qubits~\cite{Chiaverini04, Reed12, Barends14, Nigg14, Corcoles15, Kelly15,Takita16} of the system. To this end quantum error-correcting codes~\cite{Shor95, Steane96, Kitaev03, Dennis02, Terhal15, Brown14, Campbell17} have been developed to allow us to correct for a finite density of physical errors that might otherwise irreversibly disturb a quantum computation. Of particular interest are topological codes~\cite{Kitaev03, Dennis02}, which are experimentally viable due to their local structure and low-weight syndrome measurement operations ~\cite{DiVincenzo09, Wosnitzka15}.

Quantum error-correcting codes are commonly studied with a simple error model where it is assumed that each physical qubit is subject to some rate of local noise that is independent of the state of other nearby qubits and its local environment. Identifying thresholds~\cite{Dennis02} and finding the resource overheads~\cite{Bravyi13, Watson12, Brown15} of different codes using simple error models can provide a proof of principle that a code can withstand local noise, and offers a standardised benchmark to uniformly compare different error-correcting codes with one another~\cite{Landahl11, Katzgraber13}. However, when we consider some realisation of a quantum error-correcting code, where we take into account physical features of its underlying hardware~\cite{Darmawan17, Bravyi17, Iyer17}, an independent noise model is unlikely to be realistic. Instead, we might anticipate that correlated errors~\cite{Aharonov06, Ng09, Preskill13, Jouzdani14, Hutter14a, Fowler14,Novais17} could occur, where the probability of a qubit suffering an error depends on its neighbouring qubits having also suffered errors.

Correlated noise has been identified in a number of physical systems that might serve to realise quantum error-correcting codes. For example, external fields and proximity effects in solid state systems~\cite{Fowler14, OGorman14} introduce correlated errors, as do quantum codes that are coupled to a boson environment~\cite{Alicki09, Chesi10, Novais13, Jouzdani13, Freeman14, Brown14, Novais17, Delgado17}. Correlated errors have also been identified when non-Markovian effects are modelled~\cite{Ng09, McCutcheon14}. It has also been seen that correlated errors can be introduced by certain schemes of quantum error-correction. This has been observed, for instance, in Ref.~\cite{Brown15}. Correlated errors also occur when considering syndrome measurements under the circuit-based noise model~\cite{Raussendorf06, Fowler09, Fowler12b, Nickerson13, Tomita14}, and while applying transversal logical gates to quantum error-correcting codes that are not in their code space~\cite{Bravyi15}. See also recent work.~\cite{Chubb18}.

The aim of this Manuscript is to develop numerical methods to characterise correlated error models. Specifically, we develop families of decoding algorithms, each member of which is designed to correct for a particular correlated error model. We show that we can use a decoder family, which we will refer to as an adaptive decoder, to analyse a source of noise with some unknown parameter by comparing the logical error rates of different members of a decoder family. We perform simulations to demonstrate our methods explicitly by adapting Edmonds' minimum-weight perfect matching algorithm~\cite{Edmonds65, Kolmogorov09} to identify string-like correlated errors introduced to the toric code~\cite{Kitaev03}. Gaining information on the structure of correlations in noise can be used as a diagnostic tool to minimize the effect of correlations at the hardware level. It may also be used to improve classical post-processing. We show that we can use our adaptive scheme to find a decoder with an improved threshold to deal with the physical error model.

There are a number of methods and results in the literature that look to determine correlations that may emerge in fault-tolerant quantum hardware. In Ref.~\cite{Combes14} Combes~{\it et al.} showed that syndrome data can be used to discriminate between different correlated noise processes acting on small quantum codes including the classical repetition code and the five-qubit code~\cite{DiVincenzo96}. Our results build on this work by examining different codes and we give numerical evidence that these ideas are applicable to degenerate codes~\cite{DiVincenzo98}. Moreover, the use of efficient decoding algorithms that we propose make our diagnostic techniques scalable to systems of large size. We point out that recently, during the preparation of the present manuscript Ref.~\cite{Huo17} showed that it is possible to monitor time-dependent parameters of uncorrelated noise models to improve threshold error rates using methods similar to ours.

Other methods have also been found to determine correlated errors that occur during the error correction process. In Ref.~\cite{Fowler12b} the authors simulate a circuit-level error model to directly calculate the likelihood that correlated errors occur while syndrome data is measured. The information obtained with this procedure is used to improve decoding schemes. Further, in Ref.~\cite{Kelly16}, drifts in experimental parameters of a small repetition code are identified and corrected by measuring changes in the rate at which error-detection events occur. See also Ref.~\cite{Spitz17} that recently became available. These works make use of local syndrome data to determine different parameters in the error model. In contrast, we make use of a decoding algorithm that uses global syndrome data to distinguish more generic correlated error models where error events may correlate over a large number of adjacent qubits. 

We also remark on previous work to specialise decoders to account for known correlated errors. It has been shown~\cite{Duclos-Cianci10, Wootton12, Fowler13, Delfosse14a, Criger17, Varsamopoulos17, Krastanov17, Baireuther17, Maskara17} that we can obtain improved thresholds by using a decoder that specifically accounts for known correlations that occur under depolarising noise. Hutter and Loss have also identified that specialised decoders can be used to correct for known correlated errors introduced by a thermal environment~\cite{Hutter14a}. The present work extends these ideas by demonstrating that we can use learned knowledge to calibrate a decoder to give improved thresholds without complete knowledge of the error model {\it a priori}.

The remainder of the Manuscript is structured as follows. In Section~\ref{Sec:Notation} we introduce the toric code and develop decoding algorithms for general correlated errors. In Section~\ref{sec:CorrelationAnalysis} we show that the generalised decoders we present can be used to find signatures that distinguish correlated errors. Section~\ref{Sec:ImprovedThresholds} shows that we can use error model data to obtain improved decoding schemes. In Sec.~\ref{Sec:Lab} we give a discussion on how we might apply our scheme to the practical development of fault-tolerant quantum hardware. We conclude in Section~\ref{Sec:Conclusion} by discussing potential directions to develop our methods in order to diagnose more general forms of correlated error.

\section{The toric code and error models}

\label{Sec:Notation}

We study correlated errors acting on the well-studied toric code model. In this Section we briefly review the toric code, and introduce the notation we use to describe correlated error events. We go on to describe the minimum-weight perfect-matching (MPWM) algorithm that we will use to analyse correlated errors.

\subsection{The toric code}

\begin{figure}
\includegraphics[width=\columnwidth]{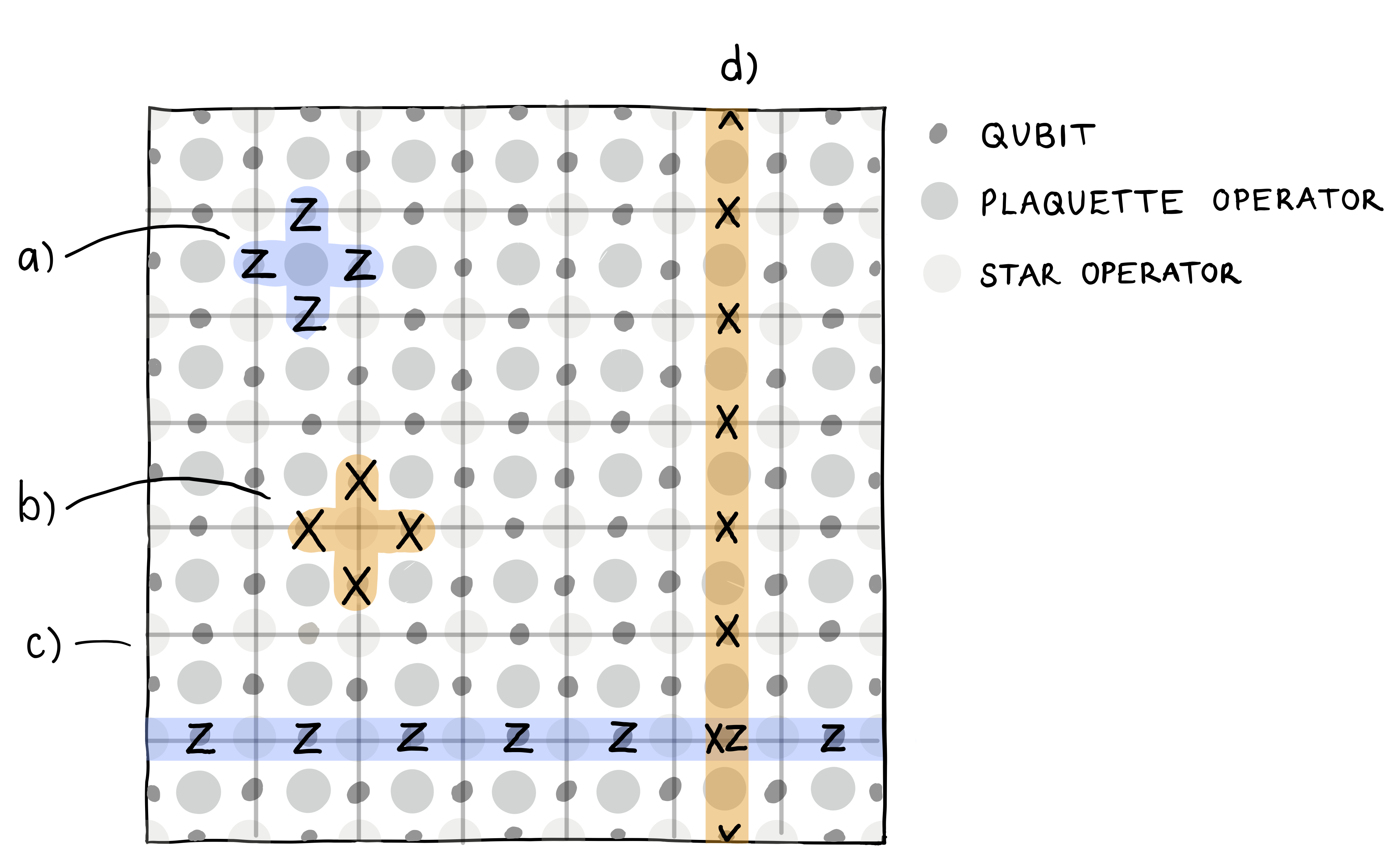}
\caption{The toric code with qubits arranged on the edges of a sqaure lattice of size $L= 7$. (a) A plaquette operator is shown in blue. (b) A star operator is shown in orange. (c) and (d) show logical operators $\overline{Z}_1$ and $\overline{X}_1$, respectively. }
\label{toric_code}
\end{figure}

The toric code~\cite{Kitaev03} is a quantum error-correcting code of $n = 2 L^2$ qubits, arranged on the edges on a square lattice of $L \times L$ vertices with periodic boundary conditions, as shown in Fig.~\ref{toric_code}. The code space of the toric code is described by its stabilizer group, $\mathcal{S}$~\cite{Gottesman97}. The stabilizer group is an Abelian subgroup of the Pauli group, $\mathcal{P}$, where the code states of the toric code, $| \psi_j \rangle$, are the common $+1$ eigenspace of all elements $S_k \in \mathcal{S}$
\begin{equation}
S_k | \psi_j \rangle = (+1) | \psi_j \rangle, \quad \forall j,k.
\end{equation}
Elements of the stabilizer group are commonly known as stabilizers.

The stabilizer group of the toric code consists of two types of stabilizers, star operators, $A_v$, and plaquette operators, $B_f$, defined as follows
\begin{equation}
A_v=\prod _{\partial j \ni v} {X_j}, \quad B_f = \prod_{j \in \partial f} {Z_j},  \label{Eqn:Stabilizers}
\end{equation}
where $j$ is an index over the qubits and $X_j$ and $Z_j$ operators are Pauli-X and Pauli-Z operators acting on qubit $j$. The set $\partial f$ are the qubits on the boundary of face $f$ and $\partial j$ is the set of vertices $v$ at either end of the edge that supports qubit $j$. The stabilizer group consists of one star operator for each vertex, $v$, and one plaquette operator for each face $f$. Colloquially, a star operator is the product of Pauli-X operators acting on all the qubits on edges incident to vertex $v$, and a plaquette operator is the product of Pauli-Z operators acting on all the edges that bound faces of the lattice, $f$. We show pictures of a plaquette and a star operator in Figs.~\ref{toric_code}(a) and~(b), respectively. 
 
 The code space of the toric code encodes two logical qubits that are acted upon by logical operators $\mathcal{L}$, whose elements are $\overline{Z}_j$ and $\overline{X}_j$ with $j = 1,2$. The logical operators are non-contractible loops of Pauli-Z or Pauli-X operators acting on the primal or dual edges of the lattice respectively. One pair of these logical operators, $\overline{X}_1$ and $\overline{Z}_1$, is shown in Figs.~\ref{toric_code}(c) and~(d). 

Quantum error-correcting codes are designed to identify and correct errors suffered by the code. We consider error operators $E \in \mathcal{P}$. The procedure for error correction involves measuring the generators of the stabilizer group. Stabilizers commute with the logical operators of the stabilizer code, and as such we do not disturb encoded logical information when stabilizer measurements are made. The outcomes of the stabilizer measurements can be used to identify the error as the elements of the stabilizer group $S_j$ that do not commute with $E$ will return a $-1$ outcome. We see this with the eigenvalue equation
\begin{equation}
S_kE | \psi_j \rangle  = - E S_k | \psi_j \rangle  = (-1)E| \psi_j \rangle, \quad \forall j,k.
\end{equation}
We refer to elements of the generating set of $\mathcal{S}$ that return $-1$ eigenvalues as defects. The collection of measurement outcomes of the selected stabilizer generators is known as the error syndrome. In the case of the toric code that we study here we use the star and plaquette operators in Eqn.~\ref{Eqn:Stabilizers} to represent the canonical set of stabilizer generators.

We require a decoder, a classical post-processing algorithm that interprets syndrome information, to estimate the most likely error that might have occurred on the lattice. A decoder will subsequently return a correction operator, $C$, to attempt to reverse error $E$. Specifically, we seek a decoder such that the value
\begin{equation}
\overline{P} = \text{prob}(CE \in \mathcal{S}),
\end{equation}
is optimised. To do so, the decoder must make use of some knowledge of the type of errors that commonly occur in the system. In the remainder of this Section we describe general correlated noise models acting on the toric code, the syndromes they give rise to, and the decoder we use to interpret the syndrome information.

\subsection{Error events}
\label{Subsec:ErrorModels}
In the previous Subsection we introduced error operators $E \in \mathcal{P}$. However, we wish to design a decoder which takes into account more information about the structure of an error operator. Here we make rigorous the concept of a structured error model by considering an error operator as a series of error events.

To study the performance of quantum error-correction procedures against correlated errors we generate random Pauli errors of the form
\begin{equation}
E = \prod_j e_j^{(1 - \sigma_j)/2}, 
\end{equation}
where operators $e_j \in \mathcal{P}$ describe a set of error events that may occur through the noise channel. Error event $e_j$ occurs with likelihood $p_j$ such that $ \sigma_j = -1 $ with probability $ p_j $ and $ \sigma_j = +1 $ otherwise.

Typically quantum error-correcting codes are tested using an independent and identically distributed(i.i.d.) noise model. For instance, the error events of the bit-flip noise model are single qubit Pauli errors acting on each qubit of the code i.e. $e_j = X_j$ for all $j$ that index the qubits of the system. Each error event in the model occurs with uniform probability $p_j = p$ for all $j$. Alternatively, the depolarising noise model is sometimes considered, where similarly, error events $e_j$ act uniformly on all qubits $j$ of the system, i.e. $p_j = p$, and then error events can take values $X_j$, $Y_j$ or $Z_j$ with uniform probability $1/3$. In our analysis that follows we will consider more general bit-flip error models with error events that are supported on multiple qubits. The restriction to bit-flip errors simplifies the problem of decoding as all of the stabilizer defects are detected by the plaquette operators.

\subsection{Decoding with the minimum-weight perfect matching algorithm}
\label{Subsec:MWPM}

\begin{figure}
\includegraphics[width=\columnwidth]{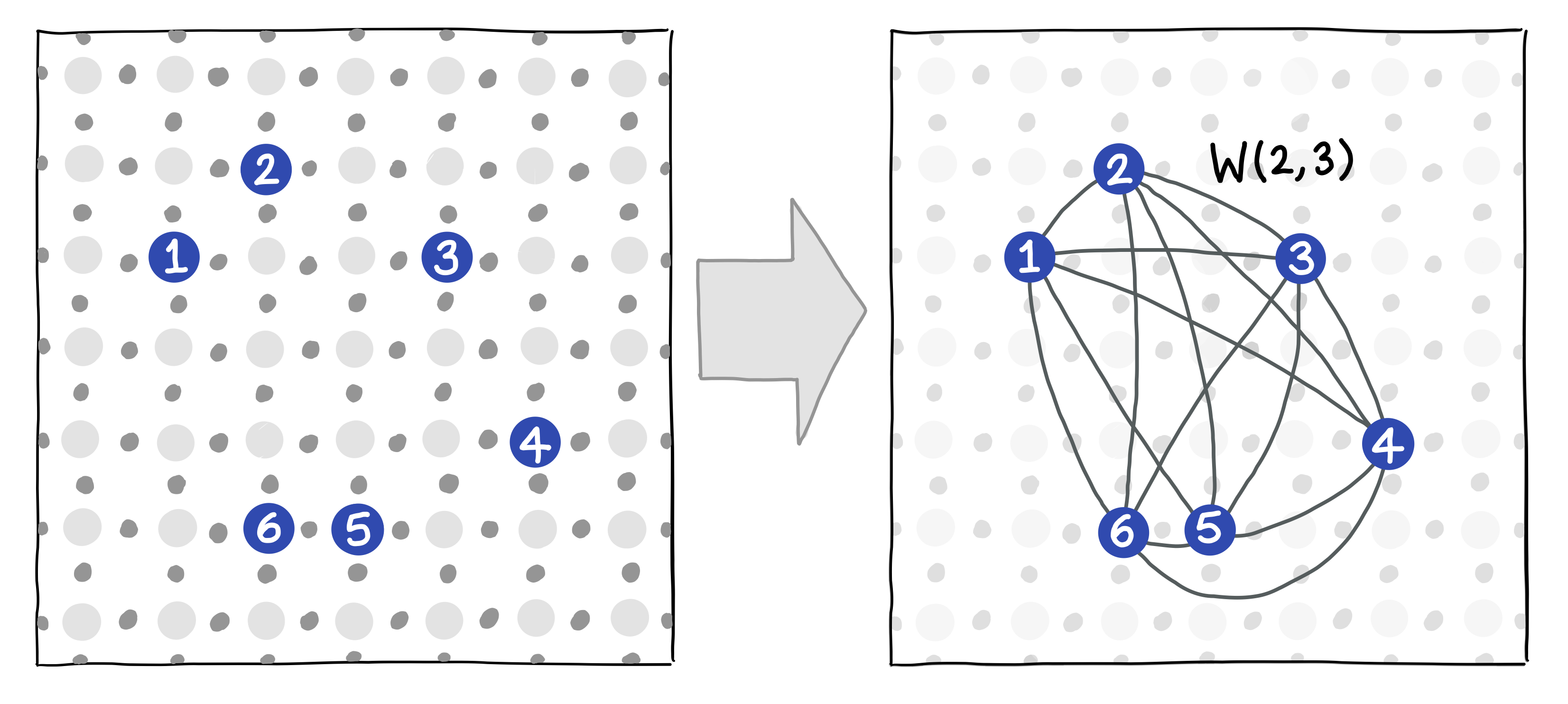}
\caption{Defects on the lattice (left) are mapped onto a graph with weighted edges (right) where the weights correspond to the probabilities that its adjacent pair of defects were created by an event that corresponds to errors along that edge.}
\label{graph_building}
\end{figure}

In the next Section we generalise the well-studied MWPM decoding algorithm~\cite{Dennis02}. For numerical studies we make use of the Blossom-V implementation of Edmonds MWPM algorithm~\cite{Edmonds65} due to Kolmogorov~\cite{Kolmogorov09}. This decoding algorithm is particularly well suited to the diffusive noise model we will introduce in the next Section.

The algorithm due to Edmonds takes as input a graph with weighted edges and returns a matching graph, i.e. a graph where each vertex has one and only one incident edge, such that the sum of the weights of all the edges are minimal. We denote edges, $e$, by pairs of vertices of the graph such that $e = (u,v)$ where $u$ and $v$ denote vertices of the graph. We also denote by $W(e)$ the weight assigned to edge $e$.

To apply the MWPM algorithm to the problem of decoding we assign to each stabilizer defect measured by a plaquette operator a vertex of the matching graph. Ideally we then construct a complete weighted graph where we weight each edge with the function
\begin{equation}
W(e) = - \log(\text{prob}(e)),
\end{equation}
where $\text{prob}(e)$ is the probability that the error model introduced an error event, or a combination of error events, such that defects $u$ and $v$ of edge $e$ occurred. In general evaluating $W(e)$ may be a complex summation of exponentially many terms. We will typically approximate this value by calculating only low-order terms of the summation. We summarise the construction of the complete graph in Fig.~\ref{graph_building}.

To find a correction operator we apply the Edmonds MWPM algorithm to the generated graph to find a matching graph where the edges correspond to a collection of error events that is likely to have caused the error syndrome. We use this information to find correction operator $C$.

We remark that for the well-studied case of i.i.d. noise we will typically have $W(e) = - \log(\text{prob}(e)) \propto d(u,v)$ where $ d(u,v) $ is the separation of vertices $u$ and $v$ of edge $e$ on the lattice~\cite{Dennis02}. Throughout this work we consider the Manhattan distance where
\begin{equation}
d(u,v) = d_x + d_y,
\end{equation}
where $d_x$($d_y$) is the separation between $u$ and $v$ along the $x$($y$) direction on the lattice.

We refer to the MWPM decoder where $W(e) \propto d(u,v)$ which is designed to deal with i.i.d. noise model as the standard decoder. Other work has also considered introducing an extra logarithmic term to account for degeneracy in errors that cause defects at vertices $u$ and $v$~\cite{Stace10, Criger17}, but for simplicity we do not use this extension here. In this work we will generalise the construction of the MWPM decoder by assigning different functions to $W(e)$ depending on the anticipated correlated noise model.

The MWPM decoder will find a single error that was likely to have caused a given syndrome with respect to an error model that is determined by $W$. We expect that our methods can be improved using maximum-likelihood decoding schemes that identify the most-likely equivalence class of errors that caused a given syndrome~\cite{Stace10, Duclos-Cianci10, Wootton12, Bravyi14}. However, in general, few efficient maximum-likelihood decoding algorithms are known~\cite{Bravyi14}. Here we assume that identifying a single error that was likely to cause the syndrome is a good approximation of maximum-likelihood decoding. This has been demonstrated to be the case for the independent and identically distributed noise model in Ref.~\cite{Dennis02, Nishimori86}.

\section{Noise characterisation}
\label{sec:CorrelationAnalysis}

In this Section we consider how to adapt the MWPM decoding algorithm, such that it can be parametrised to target different noise models. This parameterisation allows us to define a family of decoders, which we call an adaptive decoder.
Specifically, we consider a MWPM decoder where edges are assigned weights by the function $W_{\lambda}(d)$ where $d$ is the distance between the two defects, and the parameter $\lambda$ denotes different members of the decoder family. In general, $\lambda$ can represent many variables, but here, for simplicity, we will consider only integer values. We show that we can distinguish different noise models by studying logical error rates as a function of $\lambda$ with an adaptive decoder. We find that, for certain noise models with unknown parameters, the logical error rates, or thresholds of different decoders can allow us to deduce the noise parameters.

We first consider a simple example noise model to analytically justify the use of logical error rates to learn noise parameters. We will then test two different adaptive decoders, which we will define shortly, to analyse a diffusive noise model using numerical experiments.

\subsection{An illustrative example of the diagnostic tools}
\label{Subsec:BallsticNoise}

\begin{figure}
\includegraphics[width=0.7\columnwidth]{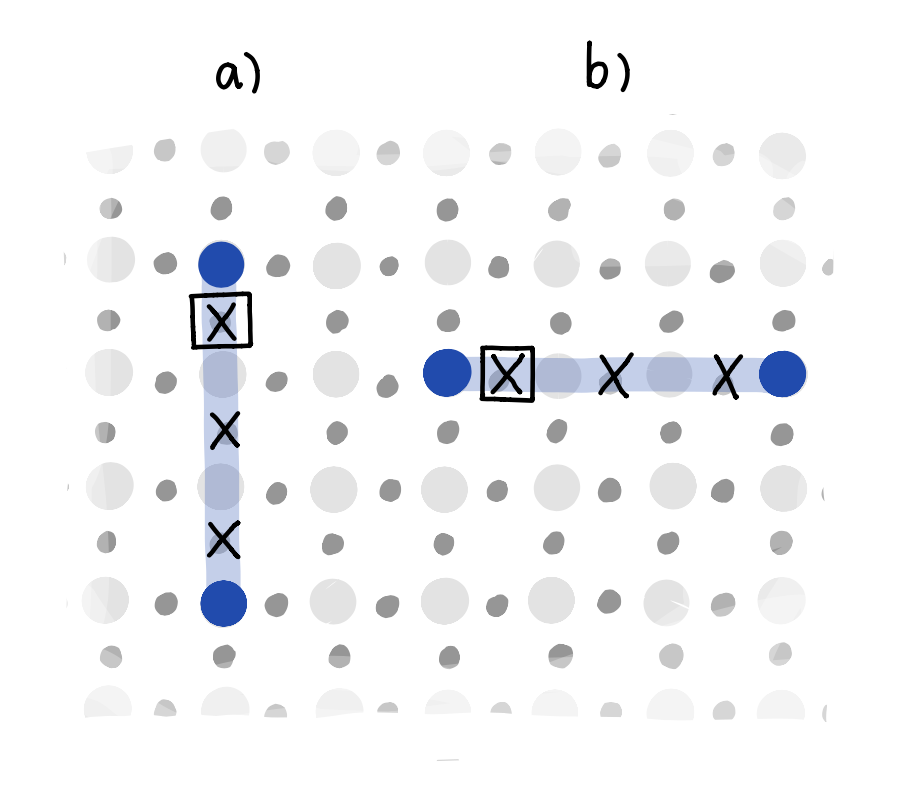}
\caption{\label{Fig:BallisticNoise} Error events of the ballistic noise model for $\xi = 3$. The model introduces error strings along vertical and horizontal lines that introduce pairs of stabilizer defects that are separated by distance $\xi$. (a)~An error event acting on a horizontal edge marked by a square box. A bit flip is introduced to the edge in the square box, and the $\xi-1$ horizontal edges immediately below the edge where the error event occurs. (b)~An error event acting on a vertical edge marked by a square box. The error event introduces a bit flip on the vertical edge and the $\xi - 1$ vertical edges to the immediate right of the edge where the event occurs. }
\end{figure}

We first develop some intuition for our diagnostic method using a simple correlated noise model, that we call the ballistic noise model, by quoting some results that are shown in Appendix~\ref{App:PathCounting}. We will argue that improved logical error rates are achieved with a targeted decoder that uses information about the noise model in the regime where the error rate is small. We use this example to motivate the simulations we conduct in the remainder of the paper. 

In the ballistic error model, error events form `straight lines' of Pauli errors with a fixed weight, $\xi$.  More specifically, the error event associated to vertical(horizontal) qubit edge $j$ introduces a bit-flip to edge $j$ and the subsequent $\xi-1$ edges that are located immediately to the right of edge $j$ (below edge $j$). Each error event occurs on edge $j$ with uniform probability $p_j = p$. We show examples of error events of the ballistic noise model where $\xi = 3$ in Fig.~\ref{Fig:BallisticNoise}. 

In Appendix~\ref{App:PathCounting} we show that for arbitrary $\xi \ge 3$ in the regime where $p$ is very low, otherwise known as the path-counting regime~\cite{Dennis02, Watson12}, the logical error rates can be improved by using a targeted decoder. It follows that we can construct a family of two different decoders to distinguish i.i.d. noise from the ballistic noise model for fixed $\xi \ge 3$ by comparing the logical error rates of the two different decoders. The two MWPM decoders of this family are the following, the  first is the standard decoder, where the matching graph is constructed with edges weighted according to the separation of stabilizer defects. This decoder seeks the lowest-weight correction operator. The second decoder in the family is the targeted decoder, which anticipates the ballistic noise model and assigns weight $a$ to edges where stabilizer defects are separated by distance $a\xi$, and otherwise assigns very high weights to edges of the matching graph.

\begin{figure}
\includegraphics[width=0.8\columnwidth]{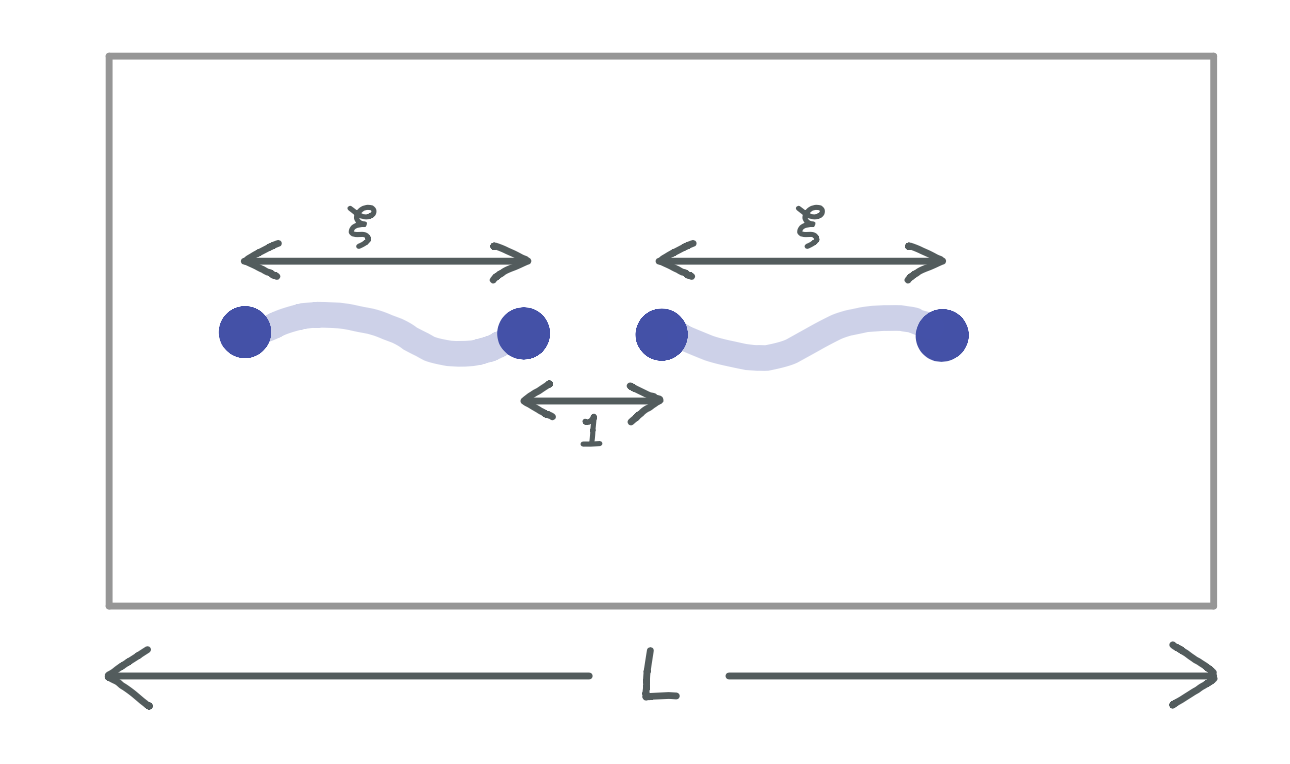}
\caption{\label{Fig:Intuition} An error caused by the ballistic noise model in the path-counting regime. The noise model introduces strings of length $\xi = L/4$. A standard decoder will have difficulty dealing with this error configuration as there are two inequivalent correction operators that have the same weight. Conversely the targeted decoder which is designed to anticipate errors that introduce defects that are separated by a distance $\xi $ will correct errors from this error model with high probability.}
\end{figure}

Consider the ballistic noise model with $\xi = L/4$ in the path-counting regime where $1/p \gg 4\xi L$. In this example the common error configurations that will cause either the standard decoder or the targeted decoder to fail are those where two non-overlapping error events occur on the same row or column. However, there are far fewer non-overlapping error configurations that will cause the targeted decoder to fail compared with the standard decoder. We give an example error configuration in Fig.~\ref{Fig:Intuition} caused by the ballistic noise model that will be corrected using the targeted decoder, but will not be corrected by the standard decoder. 

For the case of the error configuration shown in Fig.~\ref{Fig:Intuition}, the standard decoder will construct a graph that will assign equal probabilities to two inequivalent correction operators, $C_1$ and $C_2$, independent of $L$, where $C_1$ repairs the error, $C_1 E \in \mathcal{S}$, and the other correction is such that $C_2 E \not\in \mathcal{S}$. The standard decoder will struggle to deal with this case as both $C_1$ and $C_2$ will have weight $\sim L/2$ and as such, the standard decoder will not be able to identify the error events that have caused the given syndrome. In contrast, the targeted decoder that is designed to identify the error events that introduce defects separated by distance $\xi$, and will assign very low weights to the edges of the set that correspond to the correct correction operator, and very high weights to the other edges.

One can count the number of non-overlapping error configurations using Eqn.~(\ref{Eq:NumberOfConfigurations}). There are exactly $ L (L - 2\xi +1) / 2 $ error configurations of two error events where $\xi = L/4$ such that there are are exactly $L/2$ bit flips along a given row. Each of these error configurations will cause the standard decoder to fail with probability $1/2$ independent of the size of the system. However, of these non-overlapping configurations, there are only $6 \xi = 3 L / 2$ error configurations where the targeted decoder will fail with probability $1/2$, other non-overlapping error configurations of weight $L/2$ will be decoded successfully by the targeted decoder.

Having shown that there are $\mathcal{O}(L^2)$ low-weight error configurations that will cause the standard decoder to fail for the ballistic noise model, compared with only $\mathcal{O}(L)$ low weight configurations that will cause the targeted decoder to fail, we can expect to be able to identify the ballistic noise model by comparing logical error rates of the noise source with respect to the two different decoding algorithms of the family by taking system of suitably large size. In general, we show in Appendix~\ref{App:PathCounting} that $R$, the ratio of the failure probability of the targeted decoder and the failure probability of the standard decoder in the path counting regime calculated with respect to the ballistic noise model, vanishes exponentially with $L$ for constant $\xi \ge 3$. Specifically, we find
\begin{equation}
R \le \frac{ \xi + 1 }{ 2 } \left( \frac{2}{\xi} \right)^{L / 2\xi}.
\end{equation}
To this end we can expect to easily identify the ballistic noise model in the path counting regime by comparing the logical error rates of the two decoders of the family.

\subsection{Diffusive noise}
\label{Subsec:Diffusive}

We now consider a diffusive error model where each error event introduces a continuous random walk of bit-flip errors that extends across a number of nearby qubits such that two defects appear at its end points. We use this error model to numerically demonstrate our diagnostic method.

We index error events $e_f$ with the faces of the lattice $f$. Event $e_f$ will occur with probability $p_f = p$ where $p$ is uniform for all faces. Given event $e_f$ occurs, a random walk is generated for a fixed number of steps, $\xi \ll L$, between adjacent faces of the lattice. At a given step the walk direction is chosen uniformly from the four adjacent faces. Each edge, indexed $j$, that is crossed during the walk an integer $w$ times will suffer error $X_j^w$. An error event of this kind generates a single error chain, that is, where only two syndrome defects are created at the initial and final face of the walk. 

Finally, we point out that while we observe the behaviour in the case of odd and even values of $\xi$ to be qualitatively similar, we do not compare odd and even values of $\xi$ directly, as we observe there to be a slight discrepancy in their performance. We clearly see this effect in Fig.~\ref{Fig:SingleWeight}. We speculate that this is due the fact that, unlike the odd case, for even $\xi$ there is a non-zero probability that error events are members of the stabilizer group, and therefore act trivially on the code space.

\subsection{Evaluating $\xi$ for the diffusive noise model using the single-weight adaptive decoder}
\label{Subsec:FixedWeight}

\begin{figure}
\includegraphics[width=\columnwidth]{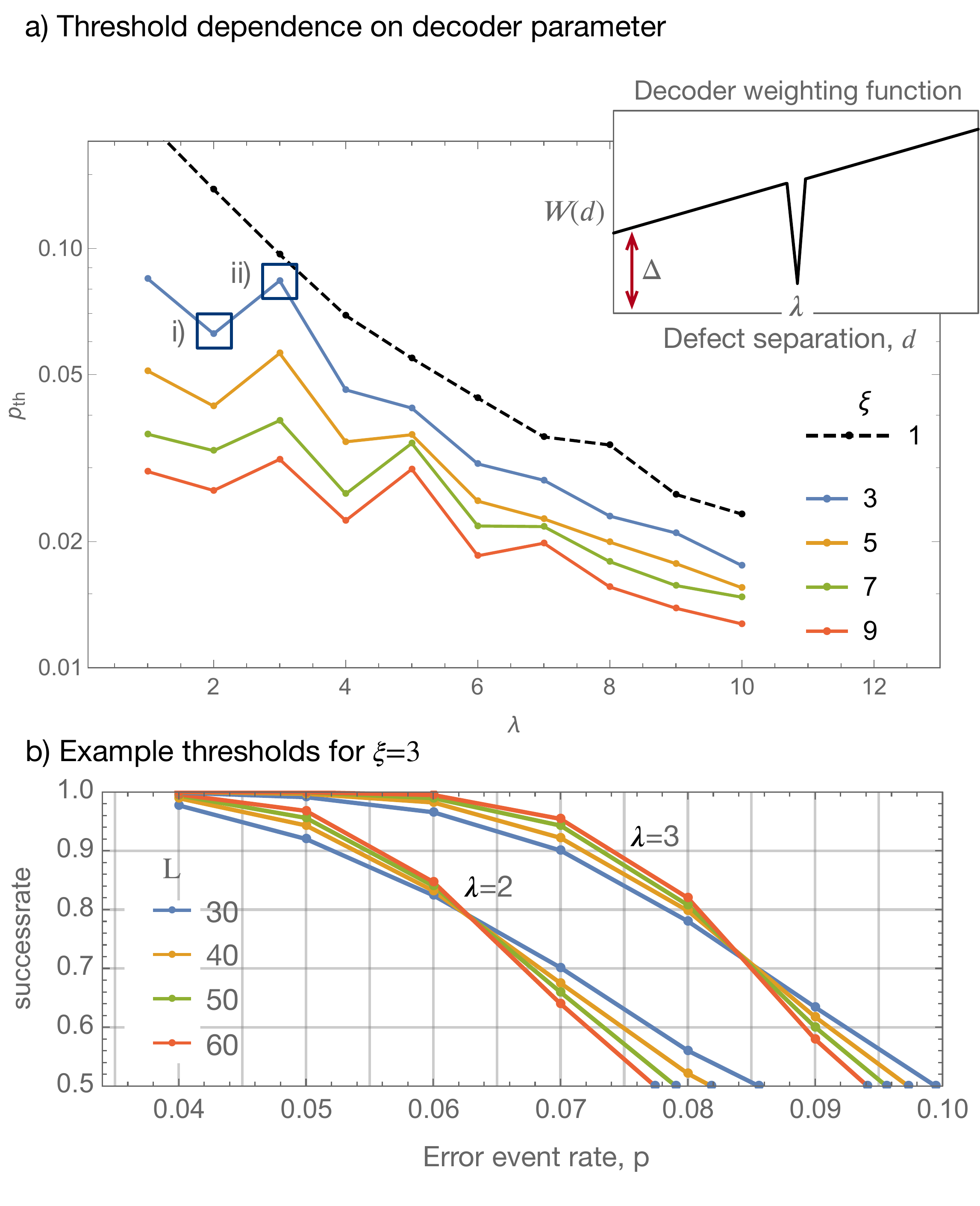}
\caption{\label{Fig:SingleWeight} 
 Threshold error rates as a function of the decoder parameter $\lambda$ under a diffusive noise model that is characterised by parameter $\xi$. (a)~Solid lines show the results for $\xi =  3, 5, 7$ and $9$. Dashed line indicates the threshold error rate for the noise model where $\xi=1$. Each value of $\xi$ shows a unique behavior. Each threshold value is obtained by repeated simulation of the noisy lattice and decoding process to identify a crossing point. Two threshold examples for a correlation parameter of $\xi=3$ are shown in (b), corresponding to the points labelled (i) and (ii) in (a).  Each point in these plots represents the aggregated result of at least 20,000 individual simulations. }
\end{figure}

Here we demonstrate numerically that using adaptive decoding the unknown parameter $\xi$ of the diffusive noise model can be identified. We use an adaptive decoder based on the MWPM decoding algorithm that we refer to as the single-weight adaptive decoder. The weighting function for this decoder assigns low weights to edges of the matching graph with length $\lambda$, and to all other edges assigns high weights that scale with the separation of pairs of stabilizer defects. Specifically, it assigns edge weights with the function
\begin{equation}
W_\lambda (d) = \left\{  
	\begin{array}{ll}
    	d 			   & d = \lambda, \\
        d\Delta \quad & \text{otherwise},
  	\end{array} \right.
\end{equation} 
where $d$ is the Manhattan distance between two defects, with $\Delta \gg L$. The weighting function for the single-weight adaptive decoder is shown { in the inset of Fig.~\ref{Fig:SingleWeight}(a)}. We will see that the logical failure rate of the single-weight decoder depends on the choice of $\lambda$ relative to $\xi$.

We numerically simulate this scenario for integer values of $\lambda$ between 1 and 10 and values of $\xi=1,3,5,7,9$. To obtain the threshold error rate we plot the logical error rate $P_{\text{fail}}$ as a function of $p$ for at least three large system sizes. To minimise the effects of simulating finite size systems we study systems where $L \gg \sqrt{\xi}$ given that we expect random walks to typically achieve walks of length $\sim \sqrt{\xi}$. The value of $p$ for which $P_{\text{fail}}$ coincides for all $L$ determines $p_{\text{th}}$. We find the crossing point by fitting the data to the function $P_{\text{fail}} = A_0 + A_1 x + A_2 x^2 $ where $x = (p - p_{\text{th}})L^{1/\mu}$ and $A_j$, $p_{\text{th}}$ and $\mu$ are constants to be evaluated. We show an example threshold plot for $\xi = 3 $ using the single-weight adaptive decoder with $\lambda = 2,\, 3$ in Fig.~\ref{Fig:SingleWeight}(b).

We remark on the clear trend in Fig.~\ref{Fig:SingleWeight}(a) where the threshold presented as a function of $p$ tend to decay quickly with $\xi$. This decay is partly due to the fact that the weight of the error increases with $\xi$ even while $p$ remains constant. A comparison of the effective single qubit error rates are shown In fact, given good knowledge of the value of $\xi$ the threshold remains relatively high even in the presence of the spatial correlations. We discuss this further in Sec.~\ref{Sec:ImprovedThresholds}.

\begin{figure}
\includegraphics[width=\columnwidth]{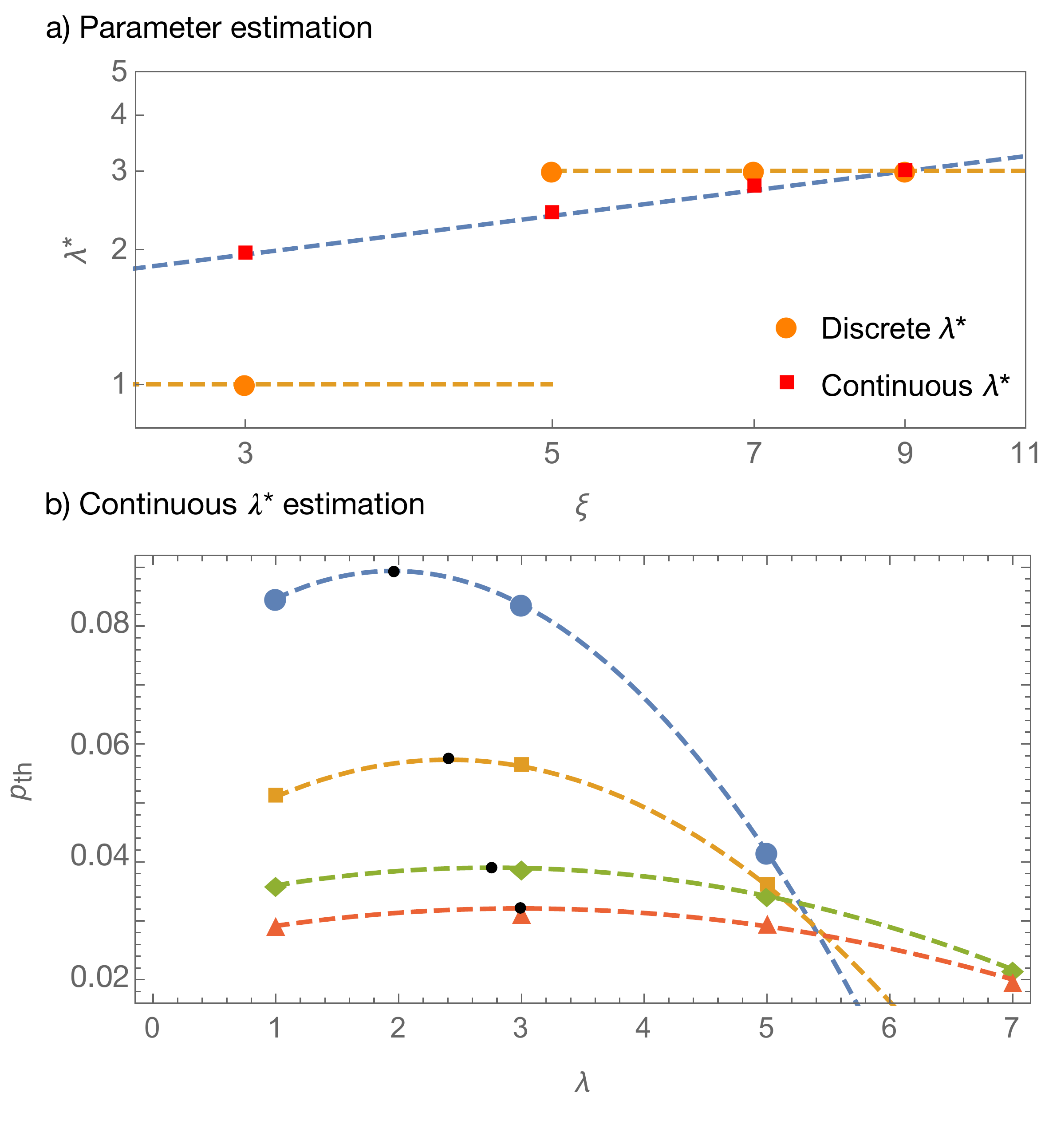}
\caption{\label{Fig:XiEstimate} Estimating $\xi$ values using signature plots in Fig.~\ref{Fig:SingleWeight}(a). (a) Estimates of $\lambda^*$ are shown. The orange points show the discrete values of $\lambda$ for which the adaptive decoder has the greatest threshold error rate. A continuous estimation is made by fitting a parabola to the data points of the signature plot with odd $\lambda$ close to the peak $\lambda$ value (red data points). This fitting is shown in (b). We observe that $\lambda^*$ grows with $\xi$ with the trend $\lambda^* \sim 1.3 \cdot \xi^{0.39}$. We show the fitting with the dotted blue line in~(a).}
\end{figure}

Having calculated the threshold for each error model using a range of decoders from the family, we go on to estimate the value of $\xi$ by plotting $p_{\text{th}}$ as a function of $\lambda$ for each noise model, as shown in Fig.~\ref{Fig:SingleWeight}(a). We observe that each value of $\xi$ generates a distinct plot as $\lambda$ is varied. We can therefore regard this as a signature of a given noise model, as such, we refer to this plot as a signature plot. We now investigate how we can use the signature plot to estimate the unknown parameter of the noise model, $\xi$ and determine other features of the noise model.

Studying Fig.~\ref{Fig:SingleWeight}(a) we first remark on the discrepancy between the data obtained for even and odd values of $\lambda$. We have considered only odd length error chains, which means single error events will never introduce defects separated by an even distance. As such, a decoder with even $\lambda$ will predominantly be correcting for second order error events, which we would expect to have a poor performance. This effect is indeed observed in the data. When the value of $\lambda$ is odd, even when it does not match $\xi$ the decoder is preferentially correcting for first-order error events. It is interesting to see that this feature of the noise model, namely that the noise model only introduces error strings of odd length, can be readily identified from the signature plot.

In Fig.~\ref{Fig:XiEstimate} we plot estimates for the value of $\lambda$ for which the threshold error rate is maximal. We denote this value $\lambda^*$. We determine $\lambda^*$ using two rudimentary methods. First of all we estimate $\lambda^*$ as the discrete value of $\lambda$ for which the threshold is maximal. We show these discrete data points in orange in the Figure. Notably, for the relatively small values of $\xi$ we study, we observe that $\lambda^*$ increases with $\xi$ in one discrete step. This observation indicates that $\lambda^*$ is sensitive to changes in the value of $\xi$.

Ideally, we would like to determine $\xi$ precisely using the signature plot. However, using the method for estimating $\lambda^*$ we have suggested so far, it is difficult to resolve $\xi$ from $\lambda^*$ as there are error models with different values of $\xi$ that return the same value $\lambda^*$. To deal with this we refine the estimate of $\lambda^*$ by using more data points from the signature plot. In the Fig.~\ref{Fig:XiEstimate}(b) we fit a quadratic curve to the odd $\lambda$ data points close to $\lambda^*$, which are shown in Fig.~\ref{Fig:SingleWeight}. The maximal values of the quadratic fittings are shown in Fig.~\ref{Fig:XiEstimate}(a) with red data points. We observe that the maximal values increase like $\lambda^* \sim 1.3 \cdot \xi^{0.39}$ which is plotted in blue against the acquired data points. We might expect a relationship $\lambda^* \sim \sqrt{\xi}$ that represents the average distance covered by an error event that performs a uniform random walk. We speculate that this discrepancy in the fitting is due to microscopic details and small size effects. Nonetheless, our numerical results show that the value of $\lambda^*$, which is found using the signature plot, estimates the unknown parameter $\xi$ precisely.

\subsection{Estimating $\xi$ using the Gaussian decoder}

\begin{figure}
\includegraphics[width=\columnwidth]{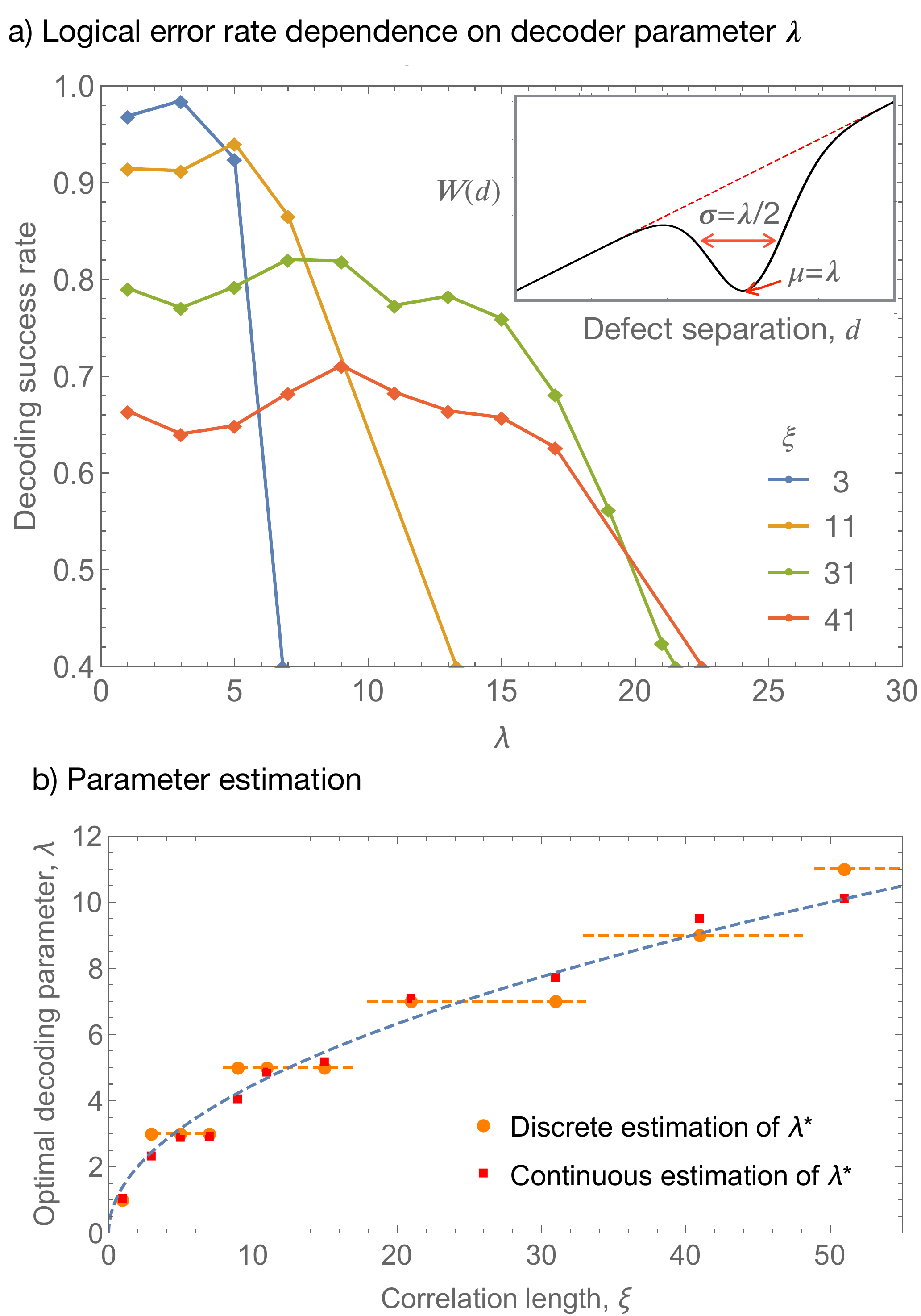}
\caption{\label{Fig:GaussianAdaptiveDecoding} (a)~Logical error rates as a function of the decoder parameter $\lambda$ are shown for values of $\xi=$3,11,31 and 41 under a diffusive noise model. The inset shows a schematic of the decoder weighting function. For each value of $\xi$ we see a peak in the logical error rate at approximately $ \lambda \sim \sqrt{\xi}$.
(b)~Estimation of the optimal decoder parameter $\lambda^*$. We identify the discrete value of $\lambda$ for which the logical error rate is maximal (orange data points), and make a continuous estimation by fitting a quadratic function around the peak (red data points). We observe that $\lambda^* \sim \sqrt{\xi}$ as we expect from a diffusive noise model where each error event will typically introduce a pair of stabilizer defects that are separated by distance $\sqrt{\xi}$.}
\end{figure}

One might argue that the adaptive decoder we used in the previous section uses too much knowledge of the error model to be applicable in any real system, since we use the fact that the random walk is of a fixed length. In this section we show that a similar result is achieved with a more `general purpose' adaptive decoder, which could apply to many more possible noise models. We call this the Gaussian decoder, as it assigns weights to the edges of the matching graph according to a Gaussian function based on the separation of two defects
\begin{equation}
W_\lambda(d) = d\left(10^4-9999\,e^{-\frac{(d-\mu)^2}{2\sigma^2}}\right)
\end{equation}
where $d$ is the separation of the two defects measured via the Manhattan distance on the lattice, $\mu=\lambda$ and $\sigma=\lambda/2$.
A schematic of the weighting function of the Gaussian adaptive decoder is shown in the inset of Fig.~\ref{Fig:GaussianAdaptiveDecoding}. The decoder is designed to assign bias to identify common error events that cause defects that are separated by a distance that is on average $d \sim \lambda$ which we expect from the diffusive noise model. Errors that are not consistent with a diffusive random walk are assigned probabilities that are proportional to their separation as we would using the standard decoder.

In the previous section we produced signature plots by computing the threshold error rate for each decoder parameter $\lambda$, and then using the result to estimate the correlation length $\xi$. In practice it will be simpler to use logical failure rates directly, rather than thresholds, as they can be computed for a single code of fixed size, and a single value of physical error rate. As such, for this case, we use logical error rates instead of threshold error rates to find a signature plot.

At the top of Fig.~\ref{Fig:GaussianAdaptiveDecoding}(a) we generate a signature plot using the logical failure rate of the adaptive decoder as a function of $\lambda$ for different odd values of $\xi$ ranging from $\xi=3$ to $\xi=51$, for a lattice size of $L=100$ such that $\sqrt{\xi} \ll L$. Only the odd values of $\lambda$ are shown. We see that, as in the previous Subsection, the peak logical success rate varies with $\xi$. In Fig.~\ref{Fig:GaussianAdaptiveDecoding}(b) we aim to estimate $\xi$ using the signature plot. Two different estimates of $\lambda^*$ are shown, a discrete estimation by choosing the value of $\lambda$ for which the logical success rate is the greatest, and an estimate based on a quadratic fit around the peak values. We observe a trend $\lambda^* \sim \sqrt{\xi}$ as we might expect from the diffusive noise model.

We finally remark that we chose a fixed relationship between $\mu$ and $\sigma$ in our weighting function, but more generally the functional form can be parametrized by more variables to explore a wider space of functions. This may be an interesting route for further work. Using our choice of adaptive decoder the results we have presented show that the logical failure rate of the decoder is sensitive to our choice of $\lambda$, and that we can use this information to improve the failure rates of our decoding scheme.

\section{Improved decoding schemes}
\label{Sec:ImprovedThresholds}

In the previous subsection we demonstrated that we can establish features of noise models by use of an adaptive decoder, and we showed this explicitly for the case of the diffusive noise model. Being able to gain a knowledge of the noise acting on underlying hardware is a valuable tool for improving the performance of quantum error-correcting codes. Particularly in the regime where the system is too large to allow for tomography. Information about the error processes involved may make it easier to identify the source of the noise, and it may be possible to improve the decoding algorithm to account for these known errors in classical post-processing. In this Section we show that we can use signature plot data to improve decoding algorithms to increase both threshold error rates and logical success rates. 

We will focus on the data obtained in Subsec.~\ref{Subsec:FixedWeight} to design better decoding algorithms. Our aim is to use signature plots to design a decoder that performs better than the best decoder we had already identified by sweeping the parameter $\lambda$. We propose a specialised decoder, that identifies the two highest peaks on the signature plot, and uses a weighting function that assigns very low weight two both of these values of $\lambda$, and high values elsewhere, otherwise the function assigns a large weight that scales with the separation of the defects. 

\begin{figure}
\includegraphics[width=\columnwidth]{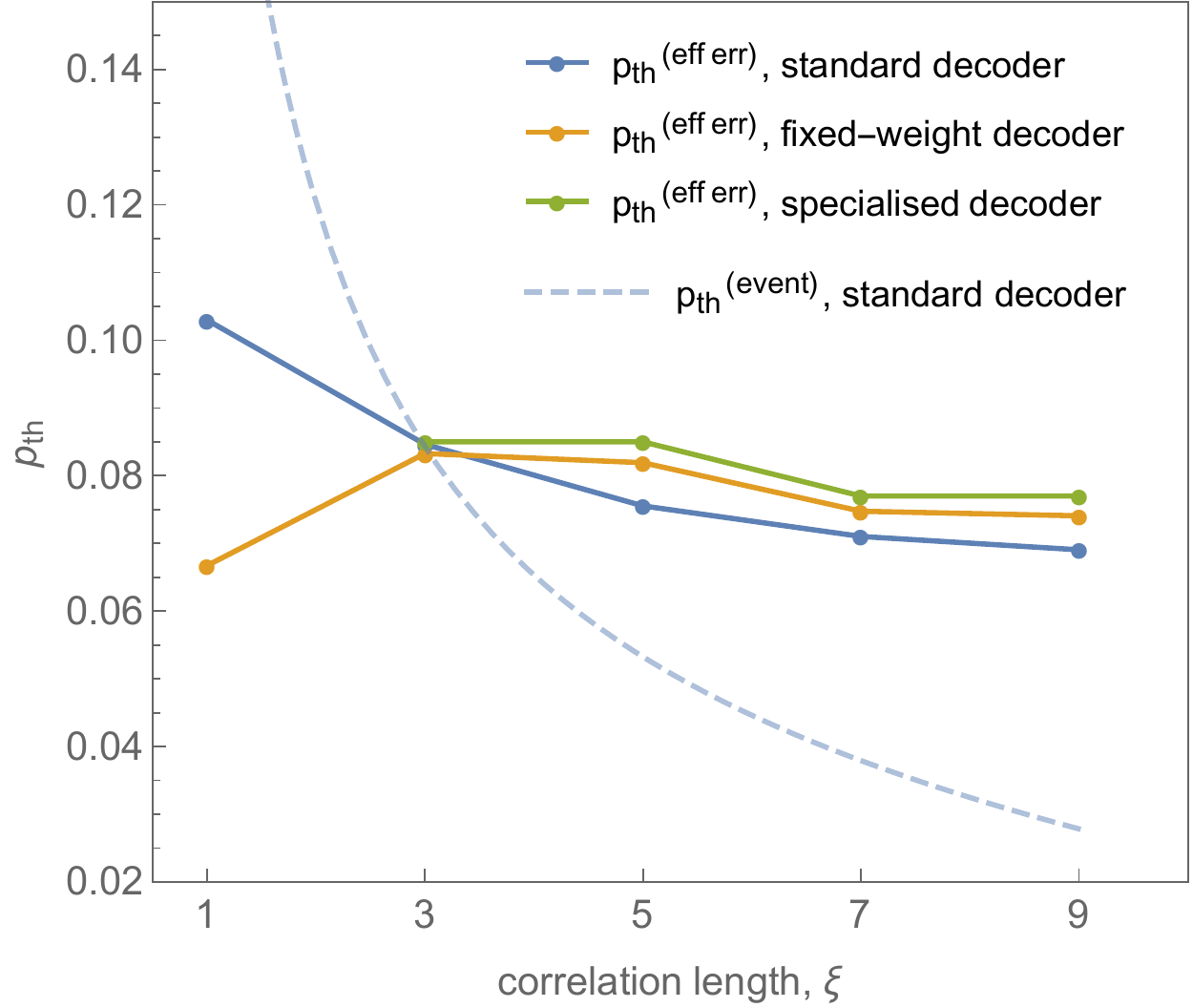}
\caption{The threshold under diffusive noise for varying values of $\xi$ for the standard decoder, the fixed weight decoder with $\lambda = \lambda^*$ and a specialised decoder where we assign a very low weight to more than one value of the $\lambda$. The threshold is shown as a function of the correlation parameter $\xi$, for the three approaches to decoding. We see that for $\xi > 3$, with an appropriate choice of a non-monotonic weighting function we obtain improved threshold error rates, thus indicating that we can design better decoders using the data available in the signature plot. The thresholds in the effective single qubit error rate are shown in the solid lines. The dashed blue line gives a comparison to the event error rate. \label{specialised_decoder}}
\end{figure}

We compare the specialised decoder with both the standard decoder, and the fixed-weight decoder which has been calibrated with $\lambda = \lambda^*$ where we use the discrete value of $\lambda^*$ found in the previous section, i.e. the orange points shown in Fig.~\ref{Fig:XiEstimate}(a). Fig.~\ref{specialised_decoder} compares the thresholds of the different decoders. We find that the specialised decoder outperforms the other decoders for sufficiently high $\xi$. The calibrated fixed-weight decoder also outperforms the standard decoder for larger values of $\xi$, but does not perform as well as the specialised decoder. Importantly, our results show that updating the weighting function using information from the signature plot improves the logical failure rates of the surface code.

We remark that the specialised decoder we have suggested here is by no means optimal, we have only demonstrated that even using a very simple method for using information from the signature plot the decoder can be improved. We expect that more sophisticated techniques will allow further improvements in the threshold, but we leave a thorough exploration of the decoder performance that can be achieved using signature plots to future work.

It is common to think about noise only as a rate of Pauli errors on single qubits, and it is useful to consider how the correlations in the noise affect this effective single qubit error threshold. Fig.~\ref{specialised_decoder} shows the thresholds in the single qubit error rate, which is defined as the probability with which any individual qubit has suffered a Pauli error at the end of the round of noise application. The effective single qubit error thresholds for the standard decoder (solid blue line) can be compared with the error event thresholds (dashed blue line). The latter decreases rapidly, since a smaller number of events introduces more single qubit Pauli errors. However, we see that even with the standard decoder, the effective error threshold decreases only gradually as correlations increase. We also plot the effective error threshold using the fixed weight decoder, and specialised decoder, as described in the previous sections. Here, we see a more gradual decline in the threshold.

It is often believed that any correlations in the noise model are highly destructive, and should be avoided. While this may sometimes be the case, for example with very long range correlations in which a single error event spans the code, the effect on the threshold depends strongly on the exact structure of the correlations. These results show that for a diffusive noise model resulting in spatial correlations much smaller than the code distance, the threshold is not strongly affected, and using a targeted decoding strategy the loss of logical error rate can be reduced.

\section{Practical considerations}
\label{Sec:Lab}

It is worth noting the practicality of the scheme we have discussed for analysis of a real experimental setup. Given that some hardware has been developed to implement an error correcting code, we can perform the following operations: prepare the code in a known logical state, perform stabilizer measurements, and measure out the logical qubit with the intention of implementing the identity gate. We can repeat this process many times to build up a large data set of measurement outcomes. A decoder can analyse the measurement outcomes to determine if a logical error occurred. We can then perform the analysis we have described in classical post-processing on this data set, using a large suite of decoders targeted at errors over a large space of possible noise models. The fact that this analysis is a classical and offline makes it a very low cost way of studying the performance of a logical qubit. Furthermore, since no direct interaction is needed with the physical device, there need be no requirement that the decoder is fast, in the sense that it can keep up with the operating speed of the physical qubits. Given that we would expect to apply our method offline, we remark that this may be a practical use case for slow decoders with high thresholds, or other useful properties, even if such decoders are impractical for online decoding.

The results of such an analysis, giving some characterization of the noise model, could then be used in multiple ways to improve a device. For example, adjustments at the hardware level may be possible to minimize sources of noise, and as we have shown, we can also improve the classical processing accompanying the error correcting code to account for the known structure of the noise.

\section{Conclusions}
\label{Sec:Conclusion}

To summarise, we have developed diagnostic tools to analyse an unknown parameter of a correlated error model acting on the toric code. Our method generalises the MWPM algorithm to define a parametrised family of decoders. We go on to compare the logical error rates of the family of decoders to hypothesise the types of correlated errors a code suffers. We have shown that a code can tolerate a higher error rate if we calibrate a decoder under the error events that commonly occur, and furthermore that an increased threshold can be found for certain spatially correlated noise models. We also discussed how our scheme could be implemented in a practical setting. 

In this work we have studied only a simple noise model, but we expect that other correlated error models where string-like errors are drawn from some distribution, such as thermal noise, will display a similar behaviour. Beyond spatial correlations, the methods we have proposed can be more broadly applied to other experimentally relevant models of correlated noise, as long as an appropriate family of decoders could be found to characterize the error. For example, errors correlated between X and Z bases, higher probabilities of error in certain physical locations, and temporal error correlations, perhaps due to malfunctions in measurement apparatus~\cite{Bombin16, Bombin14a, Nagayama17, Auger17}, are all sources of concern. Many known decoders exist which could potentially be adapted to handle these types of correlations ~\cite{Duclos-Cianci10, Wootton12, HutterWoottonLoss13, Anwar14, Hutter14, Delfosse17, Tuckett17, Maskara17}. As a first step, it will be interesting to extend the ideas we have presented here to identify multiple unknown parameters of an error model to enable us to use decoding algorithms to identify features of more realistic noise models.

Beyond the toric code, similar techniques to those we have demonstrated could potentially be used to probe string-like correlated errors in other error-correction procedures that use adaptations of the MWPM algorithm~\cite{Wang10, Delfosse14, Stephens14}. This readily extends to decoders for the toric code in the setting where measurements are not ideal~\cite{Wang03}. Moreover, it will be interesting to study local correlated error events that introduce many stabilizer defects. We expect the Metropolis-based decoders presented in Refs.~\cite{Wootton12, HutterWoottonLoss13}, neural network decoders~\cite{Torlai17}, tensor-network approximants of maximum likelihood decoders~\cite{Bravyi14, Tuckett17} or perhaps minimum-weight matching decoders that are supplemented by belief propagation~\cite{Fowler13, Delfosse14a, Criger17} may be suitable for this purpose, but we leave this for future work.

As hardware rapidly progresses towards the first implementations of small error correcting codes, it is important to consider practical methods for analysing and optimising the performance of these systems. The methods we have proposed here offer a simple way of studying and correcting for correlated errors with offline classical post processing. With further development, we hope these methods will prove a useful tool for evaluating the performance of the first generation of fault-tolerant quantum processors.

\medskip

\begin{acknowledgements}
The authors thank S. Benjamin, H. Bomb\'{i}n, D. Browne, J. Combes, M. Kastoryano, D. Poulin and J. Wootton for helpful and encouraging discussions. We are also grateful to C. Chubb comments on an earlier draft of the manuscript. We acknoweldge the use of the Imperial College Research Computing Service,  DOI: 10.14469/hpc/2232.  NHN is supported by the Engineering and Physical Sciences Research Council. BJB is supported by the Villum Foundation, the University of Sydney Fellowship Programme and the Australian Research Council via the Centre of Excellence in Engineered Quantum Systems(EQUS) project number CE170100009. 
\end{acknowledgements}

\appendix

\section{Logical Error Rates for the ballistic error model in the path-counting regime}
\label{App:PathCounting}

Here we show that a targeted decoder will have better logical error rates in the path-counting regime for the ballistic noise model for $\xi \ge 3$. For simplicity we assume only lattice sizes $L = k \xi$ for even integers $k$.

In the path-counting regime the errors that will most frequently cause the decoder to fail are where $k / 2 = L / 2\xi$ errors align along the same row or column of the lattice in a suitable configuration. The standard decoder will fail with probability independent of the size of the system if all of these $k/2$ error events occur in a configuration where no errors overlap. However, as discussed in the main text, we can design a targeted decoder that makes use of value $\xi$ to improve logical failure rates. 

There are $N_{\text{st.}}$ configurations of $k/2$ non-overlapping error events that will cause the standard decoder to fail where
\begin{equation}
N_{\text{st.}} =   \frac{2\xi}{\xi+1} \frac{(k(\xi+1)/ 2 ) !}{ (k / 2)! ( k \xi / 2  ) !}. \label{Eq:NumberOfConfigurations}
\end{equation}
In the case of the targeted decoder however there are only $N_{\text{sp.}}$ error configurations that will cause logical failure with high probability. This is because the targeted decoder is designed such that in the path-counting regime the decoder will fail if and only if all stabilizer defects of a given error configuration are separated from each other by distances that are all factors of $\xi$. There are only 
\begin{equation}
N_{\text{sp.}} = \frac{\xi k!}{(k/2)! (k/2)!}, \label{Eq:SpecialConfigurations}
\end{equation}
of such configurations. We can use the expressions given in Eqn.~(\ref{Eq:NumberOfConfigurations}) and Eqn.~(\ref{Eq:SpecialConfigurations}) to bound the ratio
\begin{equation}
R = \frac{N_{\text{sp.}} }{ N_{\text{st.}} } \le \frac{ \xi + 1 }{ 2 } \left( \frac{2}{\xi} \right)^{L / 2\xi},
\end{equation}
which is proportional to the ratio of the decoder failure probabilities for the targeted decoder and the standard decoder for the ballistic noise model. Clearly, for small $\xi \ge 3 $ the logical failure probability for the targeted decoder is exponentially smaller than the logical failure probability of the standard decoder when dealing with ballistic noise.

The remainder of this Appendix will prove Eqn~(\ref{Eq:NumberOfConfigurations}). We consider two families of functions, the first, $N_l(t)$, denotes the number of configurations that $t$ error events of fixed length $\xi$ can be aligned on a ring of $l$ sites such that none of the $t$ error events overlap. The second, $M_l(t)$, is the number of configurations that $t$ error events of size $\xi$ can be arranged along a line of $l$ sites with open boundary conditions such that none of the error events overlap. 

We have that
\begin{equation}
N_{\text{st.}} = N_{l = L}(k/2), \quad N_l(0) = 1. \label{Eqn:Observation}
\end{equation}

We show Eqn.~(\ref{Eq:NumberOfConfigurations}) recursively by finding a relationship between $N_l(t)$ and $N_l(t-1)$. To do so, we derive some facts relating the function $N_l(t)$ to $M_l(t)$. Specifically, we have that
\begin{equation}
N_l(t) = \frac{l}{t}  M_{l-\xi}(t-1). \label{Eqn:Step1}
\end{equation}
We find $N_l(t)$ of Eqn.~(\ref{Eqn:Step1}) by first considering a single error event of size $\xi$ that can be placed on any of the $l$ possible positions along a one-dimensional lattice with periodic boundary conditions. The remaining $t-1$ non-overlapping error events can be placed on the vacant $l-\xi$ lattice sites that are not occupied by the first error event. We thus count $l M_{l-\xi}(t-1)$ different non-overlapping configurations. With this expression, each error configuration counted by $N_l(t)$ is counted $t$ times. The right-hand side of the expression thus includes an additional factor of $1/t$ to account for the degeneracy. We also require the relationship
\begin{equation}
M_l(t) = N_l(t) - (\xi - 1) M_{l-\xi}(t-1). \label{Eqn:Step2}
\end{equation}
This expression is found by realising that the number of configurations included in  $N_l(t) $ that are not included in $M_l(t) $ are those where a single error event crosses one particular fixed point of the ring of $l$ sites. There are $\xi - 1$ positions that one error event may lie to cross this particular fixed point, and the remaining $t-1$ non-overlapping error events may lie in any of $ M_{l-\xi}(t-1)$ configurations. We substitute Eqn.~(\ref{Eqn:Step2}) into Eqn.~(\ref{Eqn:Step1}) to find
\begin{equation}
M_l(t) =  \left( 1 -  \frac{ t(\xi - 1)}{l} \right) N_l(t). \label{Eqn:Step3}
\end{equation}
We finally combine Eqn.~(\ref{Eqn:Step1}) and Eqn.~(\ref{Eqn:Step3}) to obtain
\begin{equation}
N_l(t) = \frac{l}{t} \left( 1 - \frac{ (t-1) (\xi-1) }{l - \xi }  \right) N_{l - \xi} ( t-1 ).  \label{Eqn:FinalResult}
\end{equation}
We recursively apply Eqn.~(\ref{Eqn:FinalResult}) to Eqn.~(\ref{Eqn:Observation}) to obtain the result of Eqn.~(\ref{Eq:NumberOfConfigurations}).



\end{document}